\documentclass[sn-mathphys,Numbered,pdflatex]{sn-jnl}

\usepackage{graphicx}%
\usepackage{multirow}%
\usepackage{amsmath,amssymb,amsfonts}%
\usepackage{amsthm}%
\usepackage{mathrsfs}%
\usepackage[title]{appendix}%
\usepackage{xcolor}%
\usepackage{textcomp}%
\usepackage{manyfoot}%
\usepackage{booktabs}%
\usepackage{algorithm2e}
\RestyleAlgo{ruled}
\usepackage{algpseudocode}%
\usepackage{listings}%
\usepackage{subcaption}

\usepackage{rotating}
\usepackage{hyperref}

\begin{document}
	
	\title[Parameter Identification by Deep Learning of a Material Model for Granular Media]{Parameter Identification by Deep Learning of a Material Model for Granular Media}
	
	
	\author*[1]{\fnm{Derick} \sur{Nganyu Tanyu}}\email{derick@uni-bremen.de}
	
	\author[2]{\fnm{Isabel} \sur{Michel}}
	\author[1]{\fnm{Andreas} \sur{Rademacher}}
	\author[2]{\fnm{J{\"o}rg} \sur{Kuhnert}}
	\author[1]{\fnm{Peter} \sur{Maass}}
	
	\affil[1]{\orgdiv{Centre for Industrial Mathematics (ZeTeM)}, \orgname{University of Bremen}, \orgaddress{\street{Bibliothekstrasse 5}, \city{Bremen}, \postcode{28359}, \state{Bremen}, \country{Germany}}}
	
	\affil[2]{\orgname{Fraunhofer Institute for Industrial Mathematics ITWM}, \orgaddress{\street{Fraunhofer-Platz 1}, \city{Kaiserslautern}, \postcode{67663}, \state{Rhineland-Palatinate}, \country{Germany}}}

	
	\abstract{Classical physical modelling with associated numerical simulation (model-based), and prognostic methods based on the analysis of large amounts of data (data-driven) are the two most common methods used for the mapping of complex physical processes. In recent years, the efficient combination of these approaches has become increasingly important. Continuum mechanics in the core consists of conservation equations that -- in addition to the always necessary specification of the process conditions -- can be supplemented by phenomenological material models. The latter are an idealized image of the specific material behavior that can be determined experimentally, empirically, and based on a wealth of expert knowledge. The more complex the material, the more difficult the calibration is. This situation forms the starting point for this work's hybrid data-driven and model-based approach for mapping a complex physical process in continuum mechanics. Specifically, we use data generated from a classical physical model by the MESHFREE software \cite{MESHFREEHomepage} to train a Principal Component Analysis-based neural network (PCA-NN) for the task of parameter identification of the material model parameters. The obtained results highlight the potential of deep-learning-based hybrid models for determining parameters, which are the key to characterizing materials occurring naturally, and their use in industrial applications (e.g. the interaction of vehicles with sand).}

	
	
	
	\maketitle

	\section{Introduction} 
	\label{sec:intro}
	In engineering, natural sciences, and industry, partial differential equations (PDEs) are widely used to model a great variety of problems. They are a great tool for modeling and solving complex phenomena ranging from the motion of incompressible fluids to the electronic structure of materials, just to name a few. Usually, these models follow the full life cycle of products from classical simulation and optimization during the development phase to process monitoring and control during production. PDE models generally introduce some critical parameters, which have to be calibrated so that the model reflects the system or problem being considered. These parameters could be scalar or space and time-dependent parameter functions, and their calibration process usually requires multiple runs of the model. In some scenarios, one has access to the solution of the PDE or observation of the system and wishes to infer the parameters underlying the governing PDE, thus an inverse problem. A wide range of inverse problems have been studied, such as tomography \cite{baguer2020computed}, inverse kinematics \cite{tejomurtula1999inverse}, and inverse problems in signal processing \cite{del2019inverse} and even in quantum mechanics\cite{cao2022neural}. However, PDE-based inverse problems are one of the most challenging inverse problems. The complexity of PDE-based inverse problems is compounded by the fact that their solutions are typically nonlinear. This further emphasizes the need for efficient and fast solvers.
	
	While traditional or standard numerical methods such as finite differences and finite elements have been used extensively to solve PDEs, most if not all these standard PDE solvers suffer from the curse of dimensionality \cite{bellman1957dynamic}, i.e. the computational cost grows exponentially as the dimension increases. This has led to the extensive study of data-driven concepts, particularly, neural network approaches for solving PDEs over the last few years. In addition to their potential of overcoming the curse of dimensionality,  these data-driven concepts usually have the potential to complete mathematical-physical models as even the finest detail or tricky non-linearity is contained in a sufficient dataset. Also, since the parameters to be determined most often are not arbitrary, but follow an unknown, application-specific distribution, the training data provides a means to recover and exploit this distribution. 
	
	This paper looks at a PDE-based inverse problem in the field of continuum mechanics, which is applicable to the automobile development process. Specifically, our focus is on a physical model of soil over which vehicles ride. The rest of this work is structured as follows: We continue in Section \ref{sec:ROM} by looking into reduced order models (ROM) and how proper orthogonal decomposition (POD) as well as deep learning (DL) can be used in ROMs. We equally highlight in Section \ref{NN_PDEs}, how neural networks have been recently applied for PDE solutions, parametric studies, and inverse problems. We then proceed to present the defining equations of our problem in Section \ref{sec:problem_formulation} and the laboratory test setting, which provides the basis of the MESHFREE simulations \cite{MESHFREEHomepage} used for the data generation. Section \ref{sec:proposed_method} presents the method used to approach the problem, i.e. PCA-NN. In Section \ref{sec:numerical_results}, we summarize the numerical results, followed by concluding remarks in Section \ref{sec:Conclusions}.
	
	\subsection{Reduced Order Models and POD/PCA} \label{sec:ROM}
	Full-order models (FOM) like the finite difference method (FDM), finite element method (FEM), finite volume method (FVM), discontinuous Galerkin method (DGM), etc. that discretize the PDEs are usually highly accurate but very expensive. Depending on the application and the goals set, the user has to balance accuracy and computation time as an algorithm of higher accuracy implies higher computation time. In FDM, for example, a finer discretization of the domain (grid) leads to higher accuracy. The result of this is a system of linear equations with many more unknowns/parameters (i.e. the solution vector has a higher dimension); thus, a larger matrix system has to be solved to obtain the PDE solution on this fine grid. This is a major setback for real-time applications, and other settings where the PDE has to be queried multiple times. Reduced Order Models (ROM) offer a solution as they seek to reduce the dimension of the solution vector while maintaining the problem's physical features. The Reduced Basis (RB) method, which has received a lot of attention in the last decade \cite{devore2017theoretical, haasdonk2017reduced, karcher2016certified, patera2007reduced, quarteroni2015reduced, rozza2008reduced} but can be traced back to the 1980s \cite{fink1983error, noor1980reduced, porsching1987reduced}, is unarguably one of the most popular ROM. This method consists of an offline and an online stage. During the offline stage, a reduced basis is obtained from a good choice of parameters, and this is used to obtain solutions of the PDE for new parameters. This is very similar to neural operator methods for solving PDEs like Fourier Neural Operator (FNO) \cite{li2020fourier} and Deep operator network (DeepONet) \cite{lu2019deeponet}. The RB method can also be extended for parameter identification tasks \cite{liu2008rapid} as well as inverse problems \cite{garmatter2016reduced}.
	
	Recently, Deep Learning-based reduced order models (DL-ROM) have been popularized to efficiently solve PDEs \cite{fresca2021comprehensive, lee2020model}. Just like the RB method, they consist of an offline (training) phase and an online (testing) phase. The DL-ROM, though time-efficient during testing, might be very costly during training due to the high number of features or dimensions of the input and/or output -- similar to RB method. The consequence of this is usually a network with more parameters, and thus more time is needed for optimizing these parameters. A common solution that reduces the number of network parameters while maintaining  or even improving the accuracy is the proper orthogonal decomposition (POD). In the field of machine learning, this is commonly known as Principal Component Analysis (PCA), used as a technique for dimensionality reduction \cite{witten2002data}. Reduced order models constructed with both deep learning and POD are referred to in \cite{fresca2022pod} as POD-DL-ROM, where accuracy-wise, they are reported to outperform state-of-the-art POD-Galerkin ROMs during the testing stage; and efficiency-wise, they outperform DL-ROMs during the training stage.
	
	\subsection{Neural Networks and PDEs} \label{NN_PDEs}
	Neural Networks have shown interesting results in dealing with high dimensional complex PDEs \cite{han2018solving}, where they overcome the curse of dimensionality for the Schr{\"o}dinger equation and Hamilton–Jacobi–Bellman equation \cite{hutzenthaler2019multilevel}, Black–Scholes equations \cite{berner2020analysis, grohs2018proof}, and Kolmogorov equations \cite{jentzen2018proof} which arise in option pricing \cite{elbrachter2022dnn}. 
	
	The popularity of neural networks in solving PDEs probably comes from the famous Physics-informed neural networks in \cite{raissi2019physics} that use a neural network to approximate a function, i.e. the solution of the PDE for a single parameter instance. Similar works include quadratic residual networks \cite{bu2021quadratic} and Deep Ritz networks \cite{yu2017deep}. 
	
	Another class of neural networks -- probably closer in its operation to RB methods -- approximate an operator by a neural network. They are known as neural operators and can be used to query solutions of different parameter instances when trained. The PCA-based neural operator \cite{bhattacharya2020model}, FNO, DeepONet are part of this class as well as other novel methods and `variants' like the Multiwavelet-based operator \cite{gupta2021multiwavelet}, graph neural operator \cite{li2020neural}, wavelet neural operator \cite{tripura2022wavelet}, and many more. \cite{nganyu2022dlpde} provides a good overview and extends them for parametric studies as well as inverse problems.

	\section{Problem Formulation}
	\label{sec:problem_formulation}
	To shorten the design cycle of vehicles and reduce the cost of development, the automotive industry employs numerical simulation tools in the vehicle development process for testing and analysis. In this application example, we are interested in the interaction of vehicles with various roadbeds such as sand, snow, mud, etc. Vehicle stability depends largely on this interaction, and the safety of the passengers is thus a concern. To approach this problem, a full-body model of the vehicle dynamics is needed as well as proper modeling of the roadbed. Of interest to us, is the modeling of the roadbed consisting of granular material. This is a continuum mechanics problem that involves not only the well-known conservation equations of mass, momentum, and energy, but also a supplementary phenomenological material model. While the former specify the process conditions and are generally well understood, the latter relates the applied strain to the resulting stress and comes with uncertainties as well as non-linearities. Obviously, the overall goal is for the simulations to match the real-life experiments, thus the selected material model is of great importance. 
	
	\subsection{Barodesy Model} \label{sec:barodesy_model}
	Material models have parameters that are specific to the considered material as well as its reaction to external conditions, and these models range from simple to complex. By using single-parametric models for the granular material (roadbed), for example, the deviation between simulations and experiments increases as the simulation time progresses. As a result, complex material models with many more parameters are used. Such parameters are usually determined by a great wealth of expert knowledge, and costly experiments. The barodesy model \cite{kolymbas2012barodesy2, kolymbas2012barodesy1} is one of such complex material models which conforms to the basic mechanical properties of the material. It is formulated in tensorial form by Equations \eqref{eqn:barodesy_S}--\eqref{eqn:barodesy_e}
	\begin{align}
		\frac{d \mathbf{S}}{d t} &= \mathbf{W S}-\mathbf{S} \mathbf{W}+\mathbf{H}(\mathbf{S}, \mathbf{D}, e) \label{eqn:barodesy_S} \\
		\frac{d e}{d t} &= (1+e) \cdot \operatorname{tr}(\mathbf{D}) \label{eqn:barodesy_e},     
	\end{align}
	with
	\begin{align*}
		\mathbf{D} &= \frac{1}{2}\left(\nabla \mathbf{v}^{\mathrm{T}}+\left(\nabla \mathbf{v}^{\mathrm{T}}\right)^{\mathrm{T}}\right)\\
		\mathbf{W} &= \frac{1}{2}\left(\nabla \mathbf{v}^{\mathrm{T}}-\left(\nabla \mathbf{v}^{\mathrm{T}}\right)^{\mathrm{T}}\right)
	\end{align*}
	and   
	\begin{align*}
		\mathbf{H}(\mathbf{S}, \mathbf{D}, e) = h_b(\sigma) \cdot\left(f_b \mathbf{R}^0+g_b \mathbf{S}^0\right) \cdot|\mathbf{D}|,
	\end{align*}
	where
	\begin{align*}    
		\sigma &= |\mathbf{S}|=\sqrt{\operatorname{tr}\left(\mathbf{S}^2\right)} \\
		\mathbf{S}^0 &= \mathbf{S}/|\mathbf{S}|, ~ \mathbf{D}^0=\mathbf{D}/|\mathbf{D}|, ~ \mathbf{R}^0=\mathbf{R}/|\mathbf{R}|  \\
		\mathbf{R} &= \operatorname{tr}\left(\mathbf{D}^0\right) \cdot \mathbf{I}+c_1 \cdot \exp \left(c_2 \cdot \mathbf{D}^0\right)\\
		h_b &= \sigma^{c_3} \\
		f_b &= c_4 \cdot \operatorname{tr}\left(\mathbf{D}^0\right)+c_5 \cdot\left(e-e_c\right)+c_6\\
		g_b &= -c_6\\
		e_c &= \left(1+e_{c0}\right) \cdot \exp \left(\frac{\sigma^{1-c_3}}{c_4 \cdot\left(1-c_3\right)}\right)-1.
	\end{align*}
	In the above expressions: 
	\begin{itemize}
		\item[-] $\mathbf{S} \in \mathbb{R}^{3 \times 3}$ is the Cauchy stress tensor (with principal stresses $\sigma_1, \sigma_2, \sigma_3$ in axial and lateral directions),
		\item[-] $\mathbf{W}$ is the antisymmetric part of the velocity gradient, 
		\item[-] $\mathbf{D}$ is the stretching tensor (the symmetric part of the velocity gradient), 
		\item[-] $e= V_p/V_s$ is the void ratio with critical void ratio $e_c$, where $V_p$ and $V_s$ are the volume of pores and solids (grains).
		\item[-] $\mathbf{v} \in \mathbb{R}^3$ is the velocity field.
	\end{itemize}
	The non-linear function $\mathbf{H}$ introduces the material parameters $c_1, c_2, c_3, c_4, c_5, c_6,$ and $e_{c0}$ which we seek to identify via deep learning in a supervised learning task, provided the stress is known. For Hostun sand \cite{desrues2000database}, for example, $c_1 = -1.7637, c_2 = -1.0249, c_3 = 0.5517, c_4 = -1174, c_5 = -4175, c_6 = 2218, e_{c0} = 0.8703$. 
	
	\subsection{Oedometric Test}
\begin{figure}[htp]
		\centering
		\includegraphics[width=\textwidth]{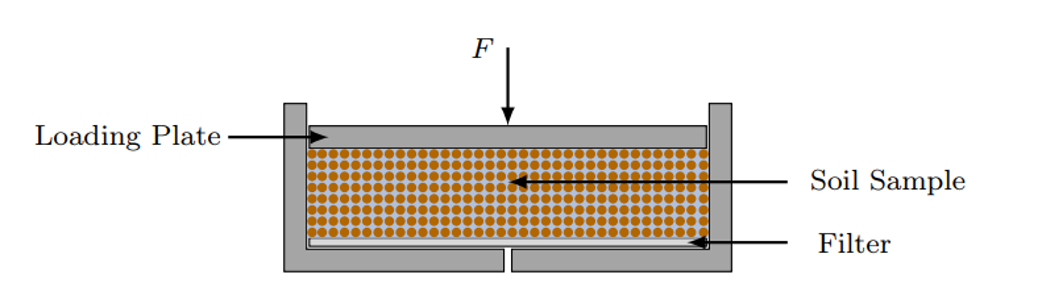}
		\caption{Schematic illustration of an oedometric test, see \cite{ostermann2013meshfree}.}
		\label{fig:oedometric_test}
	\end{figure}

	In soil mechanics, laboratory tests are used to measure the physical and mechanical properties of soil. They enable the testing and validation of material models. The tests vary from soil classification, shear strength, consolidation, and permeability tests, etc. \cite{Frattaetal2007}. The consolidation or oedometric test is one of the most conducted tests in soil mechanics. The soil (material) sample is loaded as well as unloaded in axial direction and rigid side walls prevent any lateral expansion, see Figure \ref{fig:oedometric_test}. With this, the soil's consolidation properties can be measured.
	
\begin{figure}[htp]
		\centering
		\includegraphics[width=0.6\textwidth]{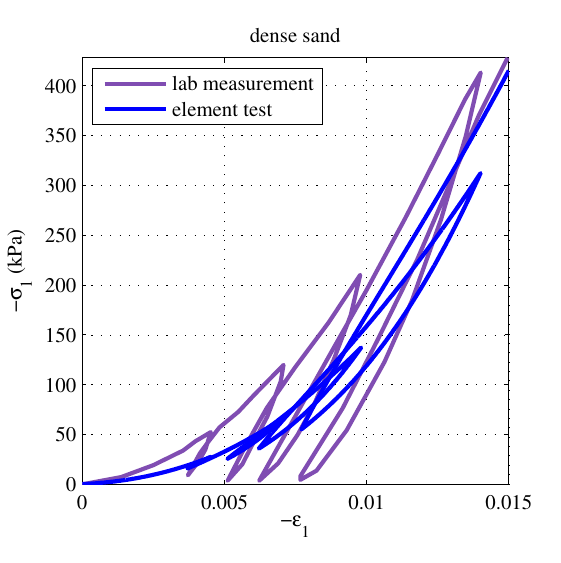}
		\caption{Axial stress-strain-curve of an oedometric test, see \cite{michel2017meshfree}, with axial stress $-\sigma_1$ and axial strain $\varepsilon_1$.}
		\label{fig:oedometric_test_stress-strain-curve}
	\end{figure}

	The laboratory measurements of oedometric tests result in stress paths (relating lateral and axial stress) and stress-strain-curves, e.g. in axial direction illustrated in Figure \ref{fig:oedometric_test_stress-strain-curve}. These are compared to corresponding element tests wrt. a material model such as barodesy, in which the material model is integrated for one numerical point. When evaluating the quality of 3D numerical methods, only the comparison with corresponding element tests should be made, since the numerics cannot be better than the material model itself. This was investigated, for example, in \cite{michel2017meshfree} for the MESHFREE software (see Section \ref{subsec:GFDM}), at that time still referred to as Finite Pointset Method (FPM).
	
	\subsection{MESHFREE and the Generalized Finite Difference Method (GFDM)}\label{subsec:GFDM}
	We employ the Generalized Finite Difference Method (GFDM) \cite{Kuhnert2014} implemented by Fraunhofer ITWM in the MESHFREE software \cite{MESHFREEHomepage, kuhnert2021meshfree} to numerically solve coupled PDEs governed by the conservation equations and material models such as the barodesy model described in Section \ref{sec:barodesy_model}. MESHFREE has successfully been applied for the simulation of complex continuum mechanics problems in industry, like vehicles traveling through water\cite{jefferies2015finite}, flow inside impulse-type turbines \cite{kuhnert2017fluid}, solution mining \cite{michel2021meshfree}, injection molding \cite{Veltmaatetal2022}, wet metal cutting \cite{Uhlmannetal2021}, and phase change processes \cite{Krausetal2023}.
	
	\subsubsection{Point Clouds and Generalized Finite Difference Approximation}
	An overview on point cloud generation for meshfree methods is given in \cite{SuchdePCGeneration2022}. MESHFREE employs an advancing front procedure \cite{michel2021meshfree} that first discretizes the boundary and then iteratively the interior of the continuum domain depending on a given point interaction radius. Each point carries the physical information (such as velocity, pressure, temperature, stress, etc.) and is moved with the continuum velocity in a Lagrangian formulation \cite{SuchdeKuhnertPCMovement2018}. Distortions caused by the movement can be corrected purely locally by adding and deleting points.
	
	Discretizing the governing PDEs in their strong formulation, GFDM generalizes classical finite differences to (scattered/irregular) point clouds. Thereby, all numerical derivatives (function values, $x$-, $y$-, $z$-derivatives or Laplacian) are computed as linear combination of neighboring function values, where the neighbors of a point are determined by the point's interaction radius. The necessary coefficients/stencils are computed by a weighted least squares method. For more details on generalized finite difference approximation we refer to \cite{kuhnert2017fluid, michel2021meshfree, SuchdePhD2018}.
	
	\subsubsection{Data Generation} \label{subsec:dataGen}
\begin{figure}[htp]
		\centering
		\includegraphics[scale=.55]{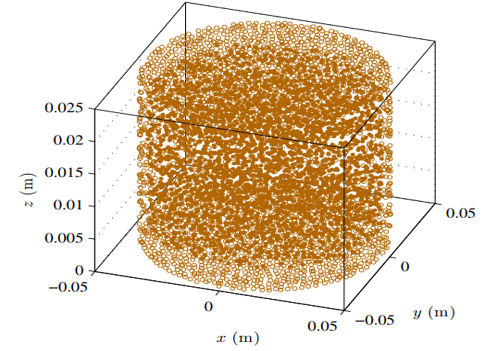}
		\caption{Initial 3D MESHFREE point cloud for the cylindrical oedometric test (filled circles: interior points, non-filled circles: boundary points), see \cite{ostermann2013meshfree}.}
		\label{fig:meshfree_t0}
	\end{figure}

\begin{figure}[htp]
		\centering
		\includegraphics[width=\textwidth]{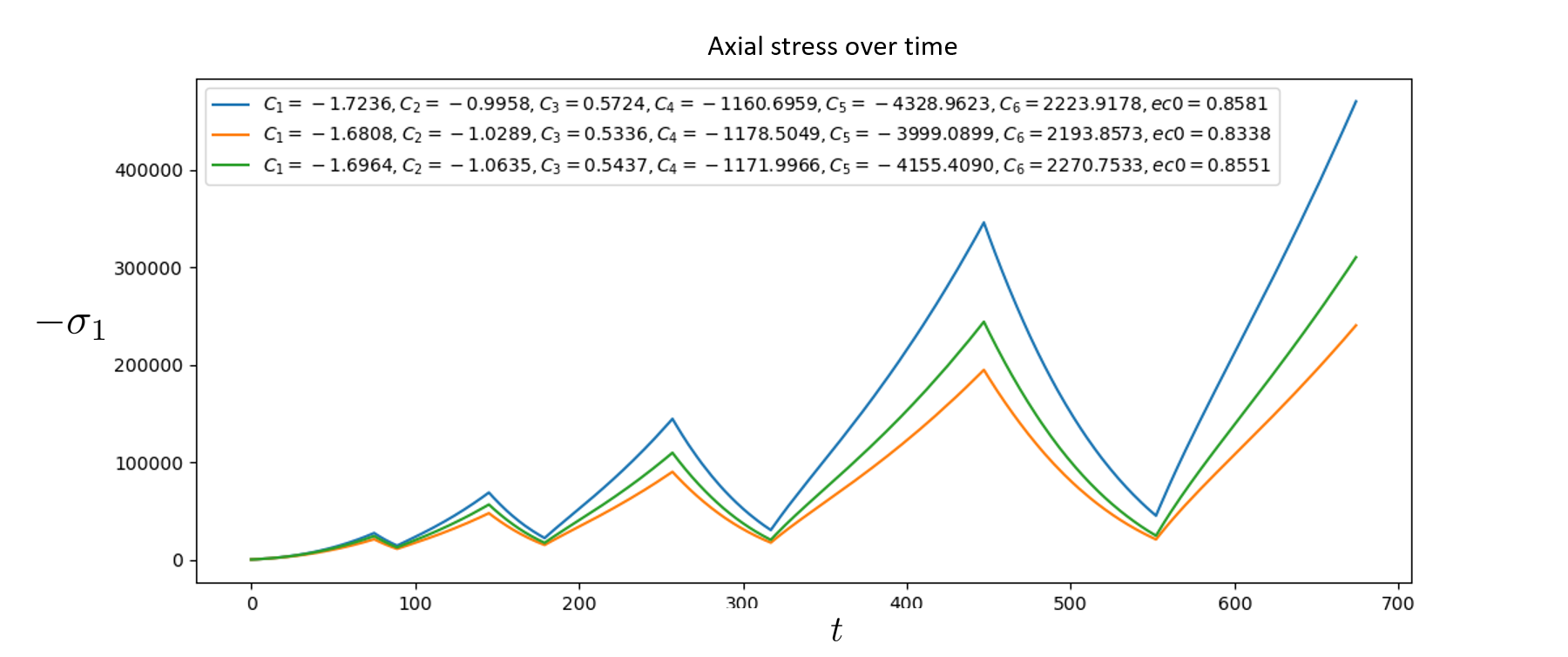}
		\caption{Example of generated input data samples for different parameters.}
		\label{fig:meshfree_data}
	\end{figure}

	Using the MESHFREE software, we generate parameters-stress pairs to train our neural network. Here, we use the physical and numerical model presented in \cite{ostermann2013meshfree} including corresponding boundary conditions for the cylindrical oedometric test. As described in \cite{michel2017meshfree}, the axial stress on the 3D point cloud (Figure \ref{fig:meshfree_t0}) is averaged over all points of the sample to determine the resulting data for a parameters-stress pair, see Figure \ref{fig:meshfree_data}. For simplicity, the representation in this figure is dependent on time and not on axial strain as in Figure \ref{fig:oedometric_test_stress-strain-curve}. Note that we use the settings for the dense sample in \cite{michel2017meshfree} with fixed interaction radius $h=0.01 \mathrm{m}$, loading/unloading rate $v_\mathrm{p}=\mp 0.001 \frac{\mathrm{m}}{\mathrm{s}}$, and fixed time step size $\Delta t = 0.0015 \mathrm{s}$ for all parameters-stress pairs.

	\begin{table}[htp]
		\caption{Value bounds for sampling of the parameters for the data generation}
		\label{tab:parameter_ranges}
		\begin{tabular}{c@{\hskip 0.3in}r@{\hskip 0.3in}r@{\hskip 0.3in}r}
			\toprule
			Parameters & Base value ($\mathbf{BV}$) & Lower bound ($\mathbf{BV}-5\%$) & Upper bound ($\mathbf{BV}+5\%$)  \\
			\midrule
			$c_1$ & $-1.7637 \times 10^0$ & $-1.8519 \times 10^0$ & $-1.6755 \times 10^0$\\
			$c_2$ & $-1.0249 \times 10^0$ & $-0.1076 \times 10^0$ & $-0.9737 \times 10^0$\\
			$c_3$ & $0.5517 \times 10^0$  & $0.5241 \times 10^0$ & $0.5793 \times 10^0$\\
			$c_4$ & $-1.1740 \times 10^3$ & $-1.2327 \times 10^3$ & $-1.1153 \times 10^3$\\
			$c_5$ & $-4.1750 \times 10^3$ & $-4.3838 \times 10^3$ & $-3.9663 \times 10^3$\\
			$c_6$ & $2.2180 \times 10^3$  & $2.1071 \times 10^3$ & $2.3289 \times 10^3$\\
			$ec_0$ & $0.8703 \times 10^0$ & $0.8268 \times 10^0$ & $0.9138 \times 10^0$\\
			\botrule
		\end{tabular}
	\end{table}
	
	The choice of the parameters that constitute the data set are selected uniformly within predefined intervals. Guided by expert knowledge (see Section \ref{sec:barodesy_model}), a base value is selected and the interval is constructed around it by adding and subtracting $5 \%$ of this base value to obtain the lower and upper bounds of this interval as shown in Table \ref{tab:parameter_ranges}.
	
	\section{Proposed Method} \label{sec:proposed_method}
	
	The proposed method is inspired by both Reduced Order Models (ROM) and Neural Networks (NN). ROMs have been popular for a long time in dealing with PDEs, and even more when dealing with parameter identification problems, as outlined in Section \ref{sec:ROM}. NNs have become popular over recent years not only due to their success in computer vision \cite{krizhevsky2017imagenet}, natural language processing \cite{hinton2012deep}, but also due to the availability of data and growing computing power \cite{goodfellow2016deep, lecun2015deep}. The efficient combination of both methods \cite{bhattacharya2020model} has already achieved remarkable results not only in simple problems but also in more complex problems such as cardiac electrophysiology \cite{fresca2020deep} (where the use of proper orthogonal decomposition (POD) further improves the results \cite{fresca2022pod}), fluid flow \cite{fresca2021real}, non-linear models \cite{cicci2022deep, fresca2022deep}, etc.
	
	\subsection{PCA-NN}
	We implement a variation of the PCA-NN architecture presented in \cite{bhattacharya2020model}, which uses a meshless operator for the evaluation of the solution of a PDE by combining ideas of ROM with deep learning. First, for given training data $(\lambda_i,u_i)$, obtain a model reduction by the use of principal component analysis (PCA) for both the input (parameter $\lambda$) and output (solution $u$). Only the coefficients of a finite number of PCA components are retained. Thus, PCA reduces the dimensions of both the input and output spaces to finite dimensional latent spaces. Second, use a NN to map the coefficients of the respective representations in these latent spaces.
	
\begin{figure}[htp]
		\centering
		\includegraphics[width=0.7\linewidth]{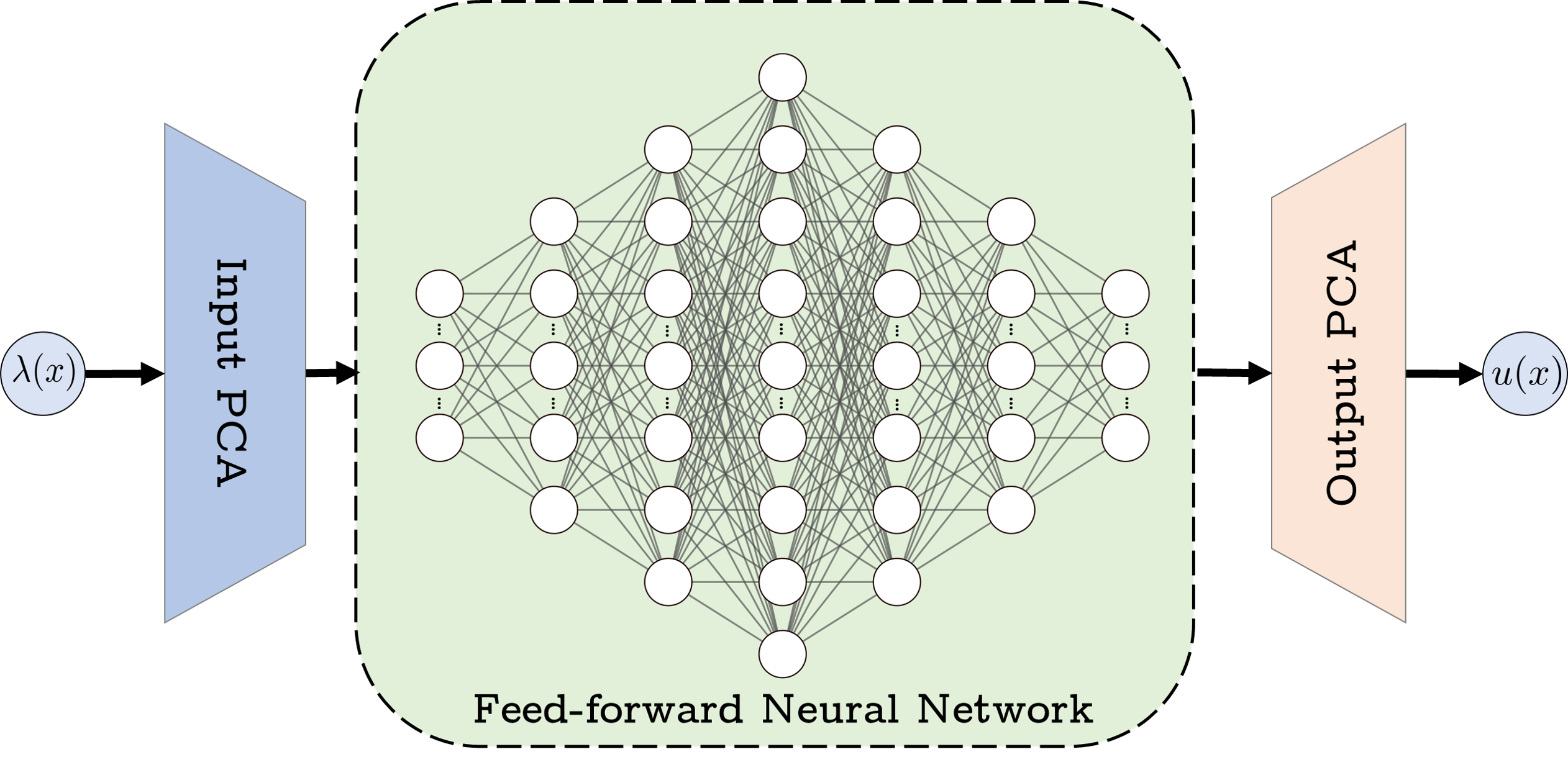}
		\caption{Architecture of the PCA-NN.}
		\label{fig:pcann_archi}
	\end{figure}

	The evaluation of this operator approximation for a novel parameter $\lambda$ is highly efficient: compute the scalar products with the specified finite number of PCA components, map these coefficients to the latent coefficients of the output space with the NN, approximate the solution of the PDE by an expansion using these coefficients and the PCA on the output side. A simplified architecture of this method is shown in Figure \ref{fig:pcann_archi}.
	
	The formulation of this approach is in a function space setting and hence mesh-free. For implementation purposes, however, we have to specify how to compute the scalar products with the PCA components. These are only given numerically, usually by their values specified at discrete points (in our case time steps).
	
	This PCA-NN operator has been used in \cite{kovachki2022multiscale, liu2022learning} in a multiscale plasticity problem to map strain to stress.
	
	\subsection{Workflow}
	In our problem, the goal is to learn the parameters  $\mu \in \mathbb{R}^{d_{\mu}}$, with $d_{\mu} = 7$, from the variation of the axial stress $-\sigma_1 \in \mathbb{R}^{d}$ over time $t$, where $d = 675$ is the fixed number of time steps corresponding to $\Delta t = 0.0015$, see Section \ref{subsec:dataGen}. The data set generated from MESHFREE is therefore a vector pair ($\mu$,$-\sigma_1$).
	
\begin{figure}[htp]
		\centering
		\includegraphics[width=\textwidth]{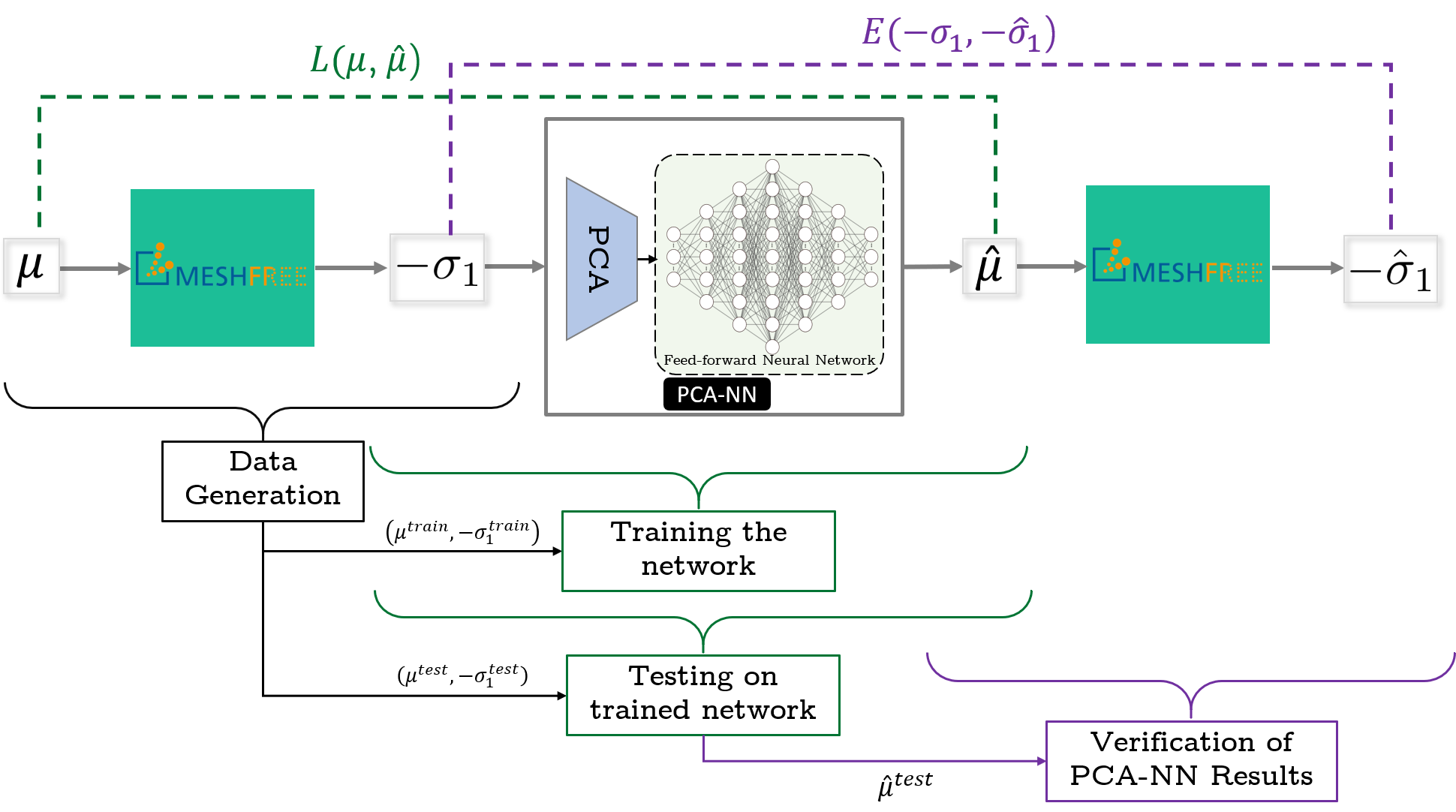}
		\caption{Complete workflow of the proposed approach.}
		\label{fig:workflow}
	\end{figure}

	Our adopted procedure can be broken down into four major steps as illustrated in Figure \ref{fig:workflow}:
	\begin{itemize}
		\item \textit{Data Generation}: Using MESHFREE and the setup described in Section \ref{subsec:dataGen}, generate parameters-stress pairs $(\mu^i, -\sigma_1^i)$ with $i = 1,2,\ldots,N_{\text{train}}+N_{\text{test}}$. These are snapshots of the full order model that is based on the GFDM described in Section \ref{subsec:dataGen}.
		
		\item \textit{Training (Offline Stage)}: The first $N_{\text{train}}$ data pairs are used to train the PCA-NN neural network. During training, the average $L^2$-loss 
		\begin{align}
			L_i (\mu, \hat{\mu}) = \dfrac{1}{d_{\mu}} \sum_{\ell=1}^{d_{\mu}}\left\|\dfrac{\mu^i_{\ell} - \ \hat{\mu}^i_{\ell}}{\mu^i_{\ell}}\right\|_{2}
			\label{eqn:loss_i}
		\end{align}
		is obtained and its average over the training data
		\begin{align}
			L(\mu,\hat{\mu}) = \dfrac{1}{N_{\text{train}}} \sum_{i=1}^{N_{\textnormal{train}}} L_i(\mu, \hat{\mu})
			\label{eqn:loss}
		\end{align}
		is optimized, see Algorithm \ref{algo:PCANN} in Section \ref{subsubsec:Algorithm} for further details. $\hat{\mu}$ is the output of the model which is a composition of PCA applied on the axial stress $-\sigma_1$ followed by the neural network.
		
		\item \textit{Testing (Online Stage)}: Once the network is trained, it is used for testing with the next $N_{\text{test}}$ unseen data. Testing proceeds as shown in Algorithm \ref{algo:PCANN_test} in Section \ref{subsubsec:Algorithm}. The network's performance is evaluated with the loss function given in Equation \ref{eqn:loss}, but averaged over the $N_{\text{test}}$ parameters by
		\begin{align}
			L(\mu, \hat{\mu}) =  \dfrac{1}{N_{\text{test}}} \sum_{i=1}^{N_{\text{test}}} L_i(\mu, \hat{\mu}).
			\label{eqn:loss_test}
		\end{align}
		
		\item \textit{Verification Stage (optional)}: This stage is used to ascertain the efficiency of the proposed model. Here, the material parameters $\hat{\mu}$ learned from the neural network are used as input to MESHFREE simulations, in order to compare the resulting stress $-\hat{\sigma}_1$ with the stress $-\sigma_1$ obtained from the ground truth parameters $\mu$. The difference is measured using the relative $L^2$-error given by 
		\begin{align}
			E(-\sigma_1, -\hat{\sigma}_1) = \dfrac{1}{N_{\text{test}}} \sum_{i=1}^{N_{\text{test}}} E_i(-\sigma_1, -\hat{\sigma}_1),
			\label{eqn:error_test}
		\end{align}
		where
		\begin{align}
			E_i(-\sigma_1, -\hat{\sigma}_1) = \left\|\dfrac{\sigma_1^i - \hat{\sigma}_1^i}{ \sigma_1^i}\right\|_{2}.
			\label{eqn:error_i}
		\end{align}
	\end{itemize}
	
	\subsubsection{Network Architecture} \label{subsec:network_archi}
	In our numerical examples, we followed the outline described in \cite{bhattacharya2020model} and used a fully connected feed-forward neural network (FCN) for the mapping of the stress latent space (output of PCA on stress) to the parameters. The number of nodes per layer starts from $d, 500, 1000, 2000, 1000, 500$, and finally $d_{\mu}$ (which is the number of parameters to be learned, here $7$). $d$ is considered a hyperparameter, which has to be tuned. In our case, $d = 50$ lead to the best results. For the PCA, we use standard randomized singular vector decomposition (SVD) implementations \cite{halko2011finding, rokhlin2010randomized}. Figure \ref{fig:workflow} illustrates the overall PCA-NN architecture.
	
	\subsubsection{Algorithm}\label{subsubsec:Algorithm}
	As a purely data-driven method, no physics or PDE is needed in the training of the neural network. However, the data used to train the network is obtained from MESHFREE's GFDM for solving the underlying PDE. By training the network with these numerically-given input-output pairs, we obtain a neural operator that solves the PDE for various instances irrespective of the underlying discretization.
	
	We specify the algorithm for the continuum mechanics problem described in Section \ref{sec:barodesy_model} using the barodesy model. The training data is the pair $\left(\mu^{i},-\sigma_1^i\right)$, with each $\mu^i\in \mathbb{R}^{d_{\mu}}$ and $-\sigma_1^i \in \mathbb{R}^{d}$. Training then proceeds as in Algorithm \ref{algo:PCANN}, while testing of the trained network proceeds as in Algorithm \ref{algo:PCANN_test}. 
	
	\begin{algorithm}[htp]
		\SetNoFillComment
		\caption{Training the PCA-NN}
		\label{algo:PCANN}
		\KwIn{\\
			\qquad$\bullet$ $\left(\mu^{i}, -\sigma_1^i\right)$: training data pair with $i = 0,\ldots,N_{\textnormal{train}}$\\ 
			\qquad$\bullet$ $\tau$: learning rate
		}
		\KwResult{Basis of PCA and the trained network $\Phi_{\Theta}$}
		Initialize the network $\Phi_{\Theta}$\\
		Compute PCA of $-\sigma_1^i$, store PCA $\left(a_{k}\right)_{k=1,\ldots,d}$\\
		Scale the respective $\mu^{i}$\\
		\For{$i \leftarrow 0, \ldots, N_{\textnormal{train}} $}{
			compute $c_{k}^{i}=\langle -\sigma_1^i, a_{k}\rangle \in \mathbb{R}^{d}$
		}
		\While{\textnormal{not converged}}{ 
			\tcc{Train network}
			\For{$i \leftarrow 0, \ldots, N_{\textnormal{train}} $}{
				Compute predictions $\hat{\mu}^i_{\ell} = \Phi_{\Theta}\left(c_{k}^{i}\right)$\\
				Compute loss $L_i = \dfrac{1}{d_{\mu}} \sum_{\ell=1}^{d_{\mu}}\left\|\dfrac{\hat{\mu}^i_{\ell}-\mu^{i}_{\ell}}{\mu^{i}_{\ell}}\right\|_{2}$ 
			}
			Compute averaged loss $L = \dfrac{1}{N_{\text{train}}} \sum_{i=1}^{N_{\textnormal{train}}} L_i$\\
			Optimize $L$ using the ADAM algorithm \cite{kingma2014adam}.\\
			Update $\Theta \leftarrow \Theta - \tau \nabla_\Theta L$ 
		} 
	\end{algorithm}

	\begin{algorithm}[htp]
		\SetNoFillComment
		\caption{Testing the PCA-NN}
		\label{algo:PCANN_test}
		\KwIn{\\
			\qquad$\bullet$ Input parameter functions $-\sigma_1^i$ with $i = 0,\ldots,N_{\textnormal{test}}$\\ 
			\qquad$\bullet$ Input PCA basis $\left(a_{k}\right)_{k=1,\ldots, d}$\\
			\qquad$\bullet$ Trained network $\Phi_{\Theta}$.
		}
		\KwResult{Output solution functions $\hat{\mu}^i$, with $i = 0,\ldots,N_{\textnormal{test}}$}
		\For{$i \leftarrow 0, \ldots, N_{\textnormal{test}}$}{
			$\hat{\mu}^i = \Phi_{\Theta} \left( \langle -\sigma_1^i, a_{k}\rangle \right)$\\
		}
	\end{algorithm}
	\begin{figure}[htp]
		\centering 
		\begin{subfigure}[b]{0.433\textheight}
			\centering
			\includegraphics[width=\textwidth]{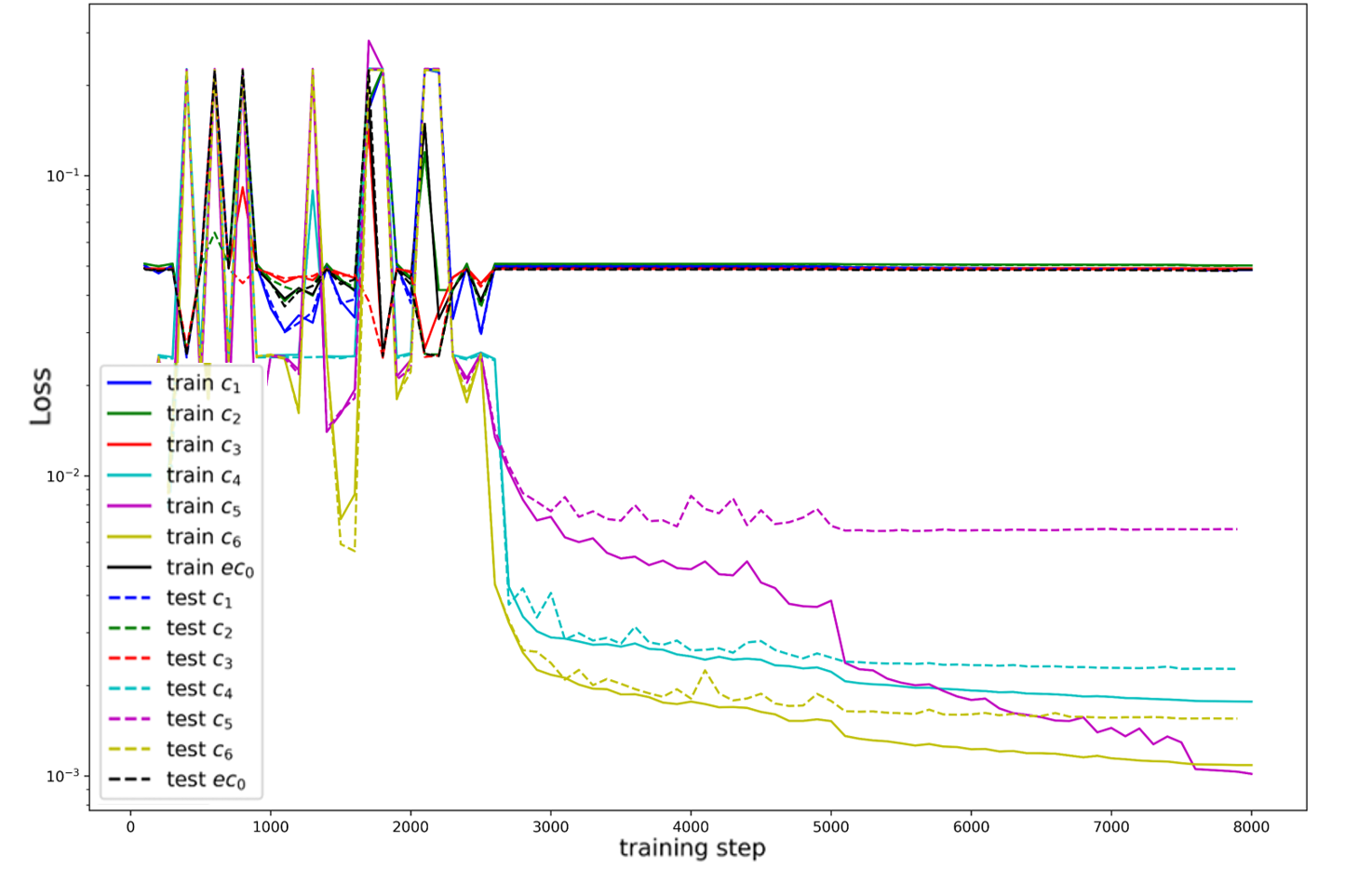}
			\caption{Individual parameters with unscaled parameters.}
			\label{fig:meshfree_loss_unscaled}
		\end{subfigure}\\
		\begin{subfigure}[b]{0.433\textheight}
			\centering
			\includegraphics[width=\textwidth]{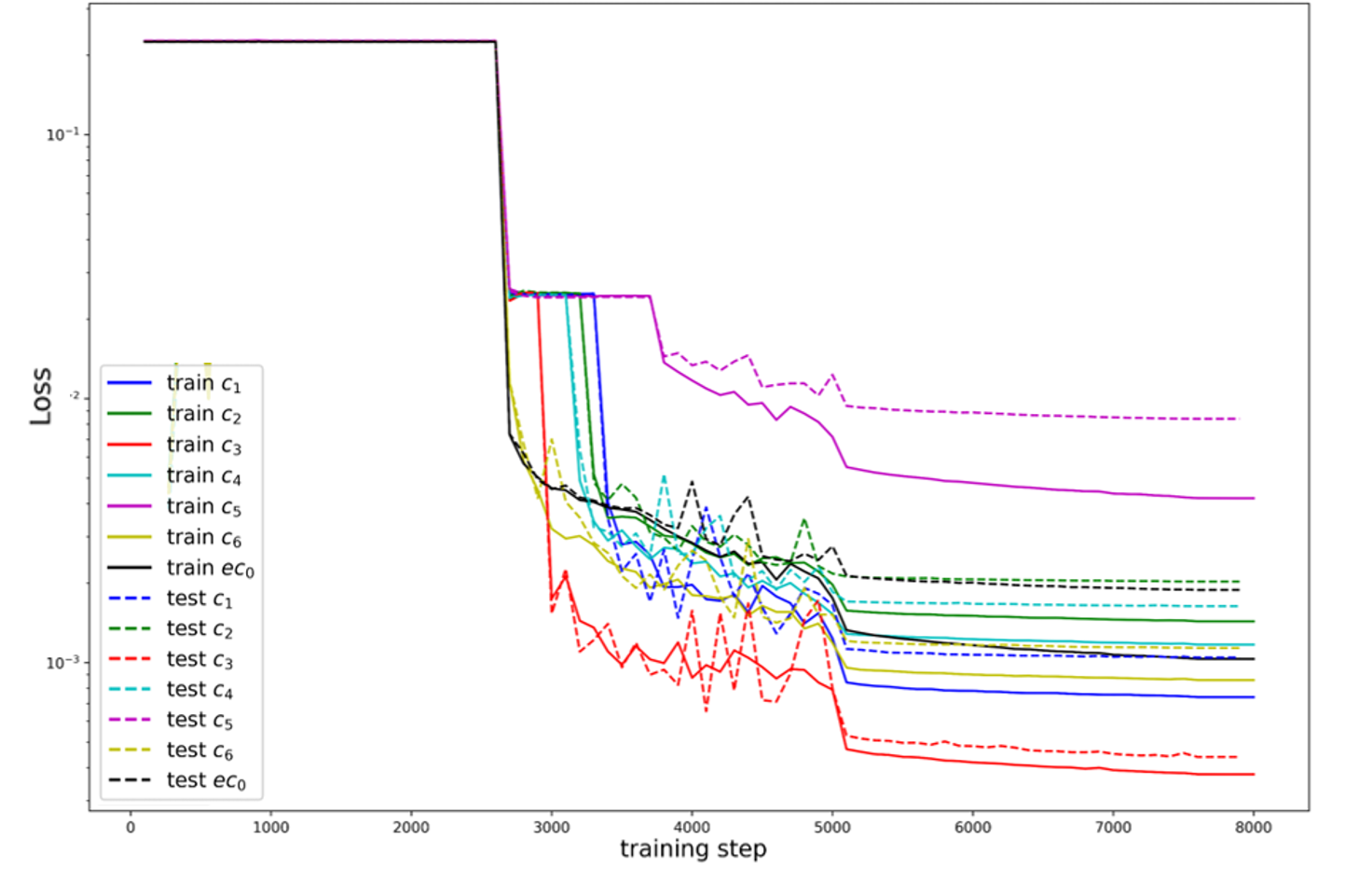}
			\caption{Individual parameters with scaled parameters.}
			\label{fig:meshfree_loss}
		\end{subfigure}\\
		\begin{subfigure}[b]{0.433\textheight}
			\centering
			\includegraphics[width=\textwidth]{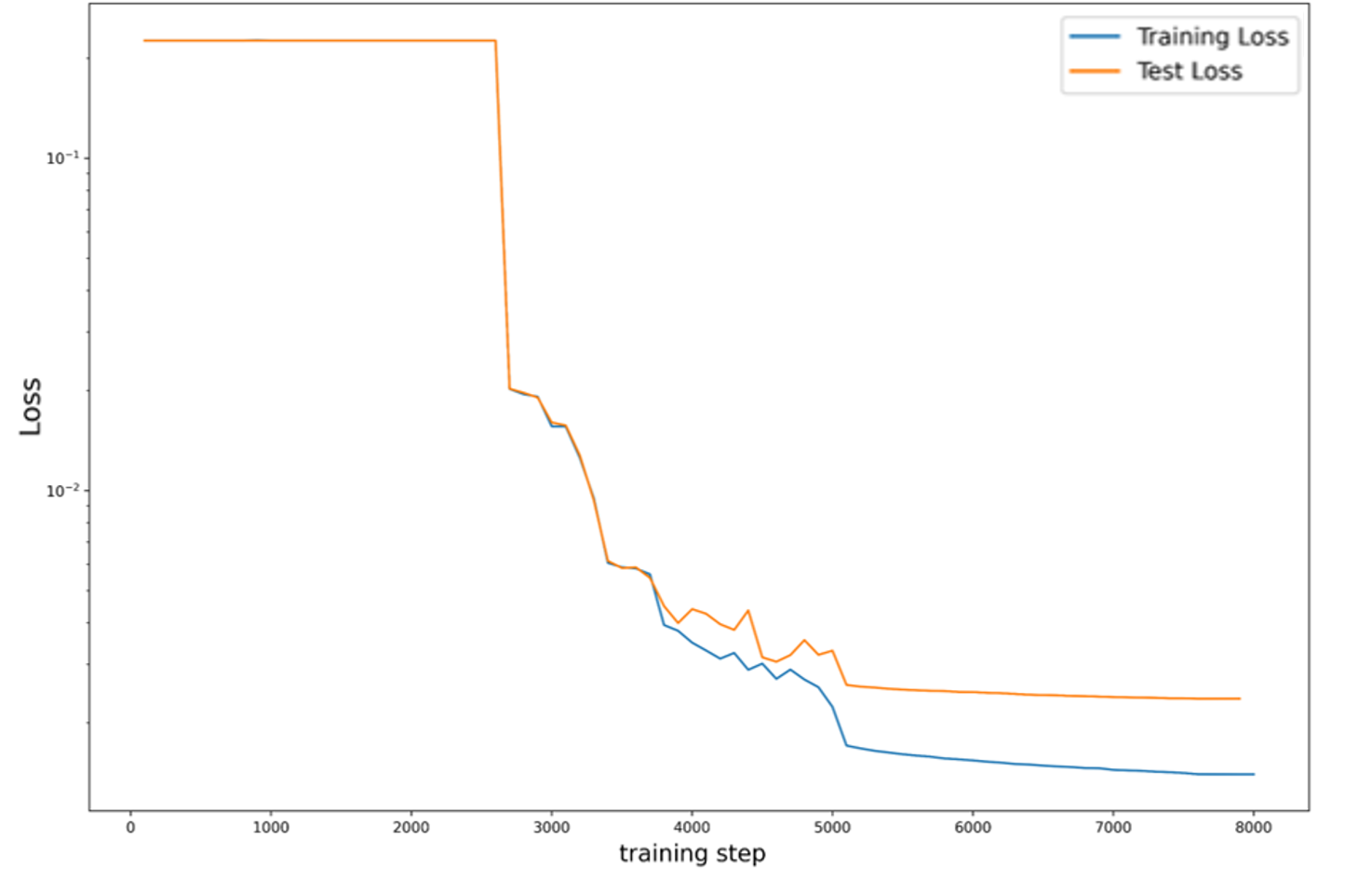}
			\caption{Average loss, with scaled parameters.}
			\label{fig:total_loss}
		\end{subfigure}
		\caption{Loss during training on both training and test data }
		\label{fig:three graphs}
	\end{figure}
	\section{Numerical Results} \label{sec:numerical_results}
	\begin{figure}[htp]
		\centering
		\includegraphics[width=\textwidth]{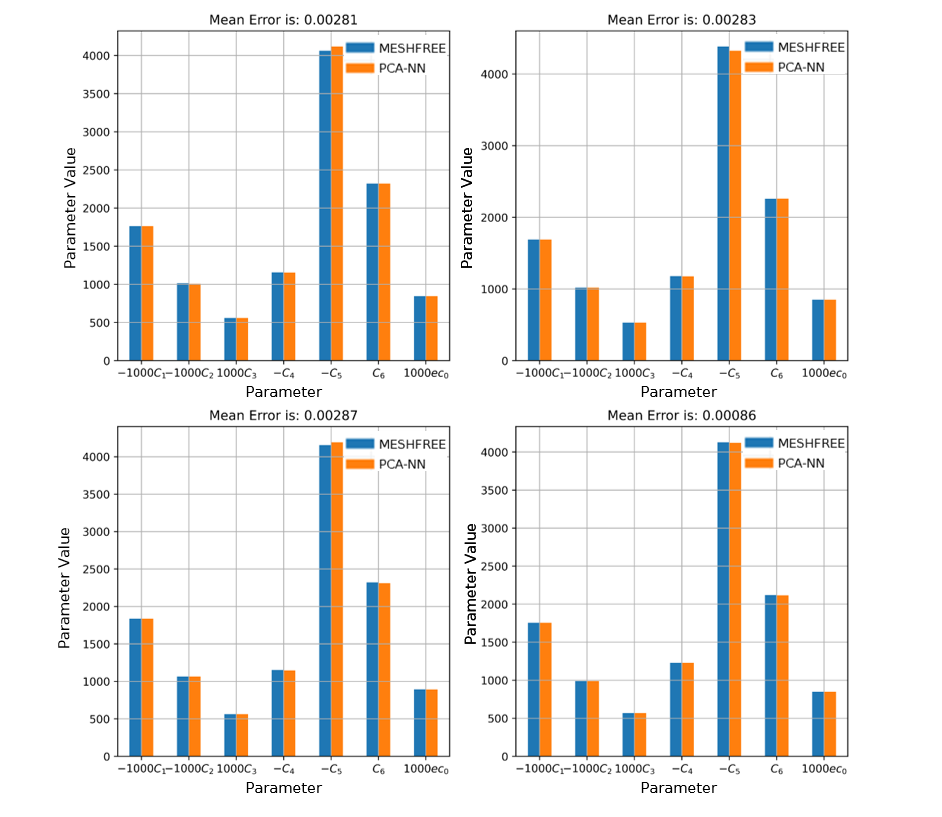}
		\caption{Comparison of ground truth (MESHFREE simulation in blue) and learned parameters (PCA-NN in orange) for four randomly chosen examples of the oedometric test.}
		\label{fig:meshfree_err}
	\end{figure}
	Randomized PCA is used to reduce the dimensions of the principal stress component from $675$ to $50$. This reduced dimension is the input to the FCN similar to that in Section \ref{sec:problem_formulation}. 
	Because the parameter space is low enough, there is no need of a PCA after the FCN. The output of the FCN yields the target parameters directly. Of the $6000$ data pairs generated, $75\%$ is used for training. During training, the relative $L^2$-error of the individual parameters is evaluated and their average is the loss function minimized for optimizing the parameters of the neural network. However, due to the nature of this loss function, the learning of the parameters of higher magnitude is favored during training as can be seen in Figure \ref{fig:meshfree_loss_unscaled}. We observe that the loss for parameters $c_4, c_5,$ and $c_6$ (that are all of the order of $1000$) is minimized, while for the other parameters (that are of the order of $1$) the loss is almost not minimized. As a remedy, the parameters of lower magnitude are scaled such that they are of the same order (of $1000$) as the parameters of higher magnitude. In this way, learning of all individual parameters is achieved as shown in Figure \ref{fig:meshfree_loss}. Figure \ref{fig:total_loss} illustrates the overall loss as average of the individual losses.
	
\begin{figure}[!t]
		\centering
		\includegraphics[width=\textwidth]{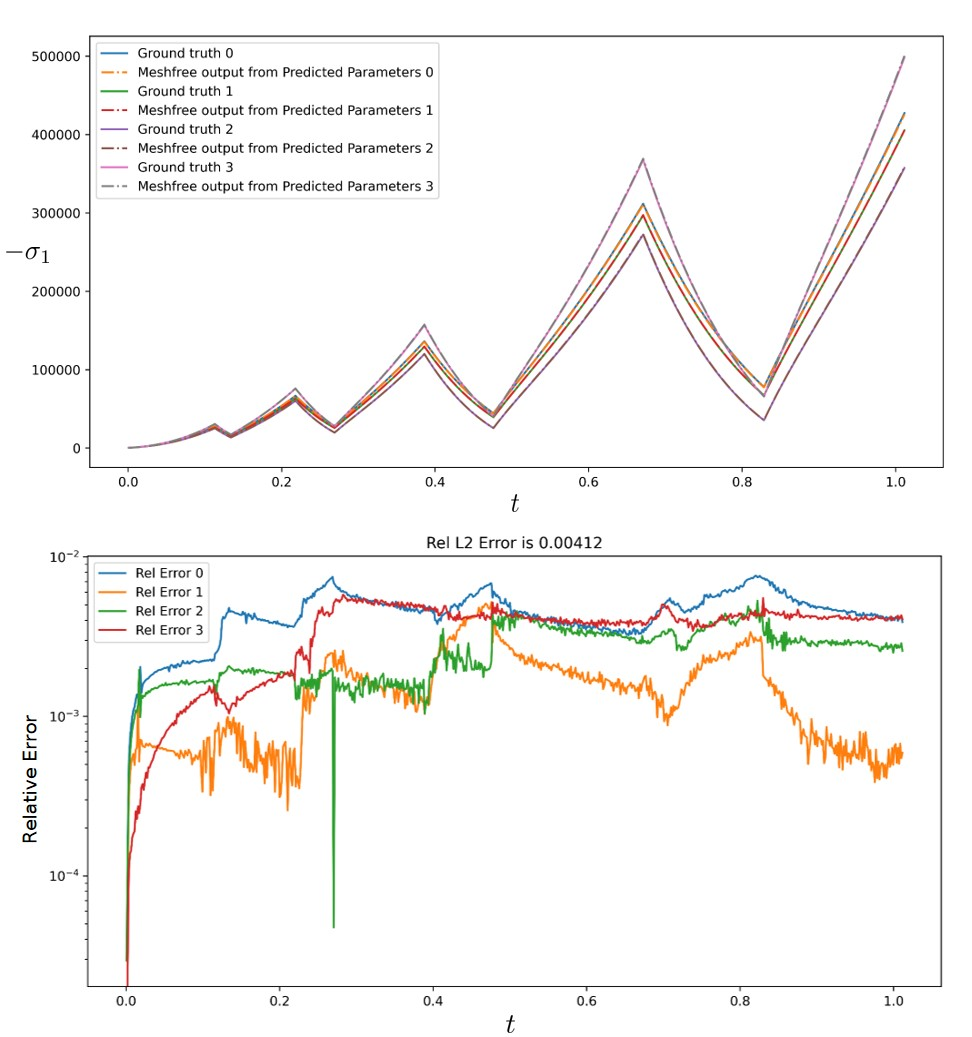}
		\caption{Comparison of MESHFREE outputs using ground truth parameters and learned parameters for four randomly chosen examples of the oedometric test (top) as well as corresponding relative $L^2$-errors. }
		\label{fig:stress_err}
	\end{figure}

	We obtained an average relative $L^2$-error of $2.63 \times 10^{-3}$ on the test data set. Figure \ref{fig:meshfree_err} shows the comparison of the ground truth (input to the MESHFREE simulation in blue) and the learned parameters (PCA-NN in orange) for four randomly selected examples. The learned parameters of these four examples were further used in a verification step in order to compare the resulting MESHFREE output axial stress with that produced by the ground truth parameters. The average relative error obtained was $4.12 \times 10^{-3}$. This is illustrated in Figure \ref{fig:stress_err} (top), where there is an obvious overlap of the axial stresses from the learned parameters with those from the ground truth parameters. Figure \ref{fig:stress_err} (bottom) shows the corresponding relative $L^2$-errors.
	
	
	\section{Conclusions and Outlook}\label{sec:Conclusions}
	
	The presented results highlight the potential of deep learning in continuum mechanics, specifically in material parameters identification for complex material models -- a task that up till now depends heavily on expert knowledge if not trial and error. By exploiting deep learning methods, we obtain the model parameters from MESHFREE simulations. It will be equally interesting to see how the results change when experimental data is used instead of or in addition to simulation data. 
	
	The proposed method is an important first step since simulation and experimental results are almost always noisy in real-life problems. An interesting future study will be to look at the effect of different noise levels on the neural network's strength in parameter identification. This is a common practice in the field of inverse problems. For example, \cite{nganyu2022dlpde} studied the effects of noise on both function-approximating networks and neural operators for PDEs. There, the PCA-based method -- when fed with noise -- did not deviate so much from the noiseless case for increasing noise level. This is also promising for our application problem.

	\bmhead{Acknowledgments}
	The authors are funded by the German Federal Ministry of Education and Research (BMBF) in the project HYDAMO. The authors would like to thank the MESHFREE team at Fraunhofer Institute for Industrial Mathematics ITWM for their support.
	
	\newpage
	\bibliography{sn-article}

\end{document}